# Uniting Weyl semimetals and semiconductors in a family of arsenides

Wojciech Pacuski

*Institute of Experimental Physics, Faculty of Physics, University of Warsaw, ul. Pasteura 5, 02-093, Warsaw, Poland*

Wojciech.Pacuski@fuw.edu.pl

**In this preview, we discuss how to combine concepts of Weyl fermions, amazing electronic properties of bulk Weyl semimetals, and advances in molecular beam epitaxy with the needs of semiconductor industry, through the fabrication of TaAs thin films.**

Since the inception of molecular beam epitaxy (MBE), arsenides such as GaAs and AlAs have been at the forefront of interest for material scientists testing basic concepts of epitaxy. The development of high electron mobility transistors (HEMTs) based on MBE-grown GaAs/AlAs interface has amplified this interest, and arsenide HEMTs have become a prominent industrial application of MBE, serving as an integral part of our everyday life by powering various devices, including mobile phones. For a long time, the primary tool to improve the performance of such electronic devices was enhancing electron mobility through better ultra-high vacuum conditions and avoiding unnecessary dopants in MBE chambers. However, there is limited room for further improvements using this approach, prompting theoretical and experimental physicists to explore various other concepts to enhance electric performance (Fig. 1).

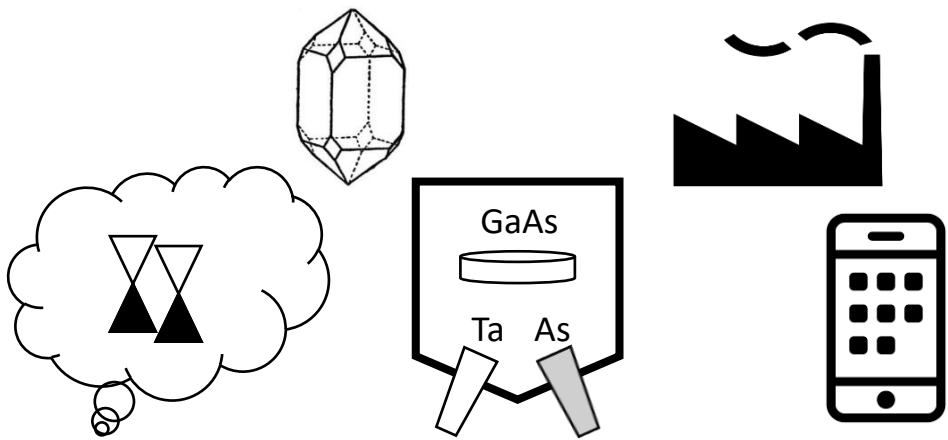

Fig. 1. We are on the way from theory of Weyl fermions, experimental demonstrations of great electronic properties of bulk TaAs crystals to more scalable methods such as molecular beam epitaxy of TaAs on GaAs substrate. Similar methods are already used for production of electronic devices.

The general strategy involves shaping the electronic band structure of materials to lower the effective mass and eventually making electrons behave like massless fermions. Moreover, leveraging symmetry and topology helps to avoid electron scattering by blocking unnecessary channels and taking advantage of surface states. While such properties are difficult to find among insulators, metals, or traditional semiconductors, they are not so rare among semimetals – materials with separated bands for electrons and holes but with zero or almost zero energy gap between them. Unfortunately, most of these fascinating semimetals, known as topological semimetals, are quite exotic materials, not very compatible with the present semiconductor industry. They are first identified theoretically and then confirmed experimentally using bulk samples. Present challenge lies in growing thin films of high quality with electron transport properties similar to theoretical predictions or properties already obtained in bulk samples.

Various difficulties arise when trying to grow such semimetals. Interesting topological semimetals usually have very different structures and chemical compositions compared to the accessible large-size, high-quality substrates used in epitaxy. Additionally, many topological semimetals require heavy metal elements, such as Cd or Hg, which are volatile and generally avoided in common technologies due to health risks. Moreover, introducing such volatile group-II elements into pure arsenide chambers poses the risk of unintentional electrical doping. Therefore, the semiconductor industry would benefit from the development of a topological semimetal system that is compatible with already existing, acceptable, and tested solutions.

This almost perfect compatibility with the present semiconductor industry is the central point of the recently published paper in Matter titled "Thin film TaAs: developing a platform for Weyl semimetal devices" by Jocienne N. Nelson et al. [1]. For semiconductor technology, the epitaxy of arsenide-based semimetals is a great opportunity to extend possibilities without any risk. Tantalum, a key component in this research, is not volatile (effective sublimation requires about 3000 C), making it safe to be used as a molecular source in any MBE chamber without the risk of unintentional doping with tantalum elements. Epitaxial growth of TaAs can be realized on standard GaAs substrates, which share a common anion – a highly favorable situation in the technology of heterostructures. TaAs belongs to the promising family of Weyl semimetals [2,3], which hold great promise for high mobility, including surface scattering-free conduction channels. Moreover, in the same family of Weyl semimetals, there are equally compatible materials such as NbAs, TaP, or NbP. Growing them in one chamber opens perspectives for Weyl semimetal alloys (e.g., (Ta,Nb)As), heterostructures (e.g., quantum wells), and superlattices of Weyl semimetals, which can further shape the electronic band structure and multiply the number of highly conductive interfaces.

For Weyl semimetals physics intensively developed last 8 years, TaAs epitaxy is a great chance to demonstrate large-scale solutions, to gain inspiration for the calculation of novel structures combining various layers, which were not accessible in bulk version. Moreover, epitaxy will also introduce several interesting challenges, such as the presence of strain, defects, proximity of surfaces, various positions of the Fermi level, and interfaces between not perfectly oriented grains.

It is worth noting that plotting a rich phase diagram illustrating the growth conditions of TaAs and demonstrating TaAs properties as a function of epitaxial layer thickness [1] required the deposition of a relatively large amount of tantalum, which is much more challenging than what one would expect from standard epitaxy of more volatile elements. For example, in the first papers about epitaxial [4] and thin polycrystalline [5] TaAs layers, which appeared less than a year ago, the available growth rates were in the order of a few nanometers per hour, and the amount of accessible Ta corresponded to 10-30 nm of TaAs. This limitation was due to the use of very thin tantalum rods, which were quickly consumed during growth. In contrast, in the case of pulsed laser deposition (PLD) techniques, the limitations were related to the amount of arsenic, which evaporates quickly from TaAs [6]. As shown in the present paper [1], thicker TaAs epilayers exhibit two orders of magnitude higher mobility and magnetoresistance, but unfortunately still not as high as in the case of bulk samples [3]. However, this is just the beginning, and there is enormous potential for inventions that can increase electron mobility.

The author acknowledges financial support from the National Science Centre (Poland) grant number 2021/41/B/ST3/04183.

The author declares no competing interests.

During the preparation of this work the author used Grammarly and ChatGPT in order to correct grammar and spelling. After using these tools, the author